\documentclass{iau} 
\usepackage{graphicx}

\newcommand{\kms}{km s$^{-1}$}
\newcommand{\kmsp}{km s$^{-1}$ pc$^{-1}$}
\newcommand{\gcloud}{G0.10$-$0.08}
\newcommand{\brick}{M0.25+0.01}
\newcommand{\am}{NH$_{3}$}
\newcommand{\meth}{CH$_3$OH}

\newcommand{\rrl}{H64$\alpha$ and H63$\alpha$}
\newcommand{\rrltwo}{H64$\alpha$+H63$\alpha$}
\newcommand{\qc}{Quintuplet cluster} 
\newcommand{\scloud}{M0.20$-$0.033}
\newcommand{\amm}{molecular gas}
\newcommand{\gc}{GC}
\newcommand{\mc}{molecular cloud}
\newcommand{\low}{25 \kms}
\newcommand{\high}{80 \kms}
\newcommand{\sick}{the Sickle H{\scriptsize II} region} 
\newcommand{\ra}{R.A.}
\newcommand{\dec}{dec}
\newcommand{\h}{$^{\mathrm{h}}$}
\newcommand{\m}{$^{\mathrm{m}}$}
\newcommand{\s}{$^{\mathrm{s}}$}
\newcommand{\degree}{$^{\circ}$}
\newcommand{\arcmin}{$'$}
\newcommand{\arcsec}{$''$}
\newcommand{\til}{$\sim$}
\newcommand{\jyb}{Jy beam$^{-1}$}
\newcommand{\pv}{position-velocity}

\newcommand{\hi}{H{\scriptsize I}}
\newcommand{\butter}{Butterfield {\it et al.} 2016, {\it in prep}}

\title[Molecular and ionized gas kinematics in the \gc~Radio Arc] 
{Molecular and ionized gas kinematics \\ in the \gc~Radio Arc}

\author[N. Butterfield, C.C. Lang, E. A. C. Mills, D. Ludovici, J. Ott, and M. R. Morris]   
{N. Butterfield$^1$,
 C.C. Lang$^1$, E. A. C. Mills$^2$, D. Ludovici$^1$, J. Ott$^3$ \and M. R. Morris$^4$}

\affiliation{$^1$Dept. of Physics \& Astronomy, University of Iowa, USA (email: {\tt natalie-butterfield@uiowa.edu}) 
$^2$Dept. of Physics \& Astronomy, San Jose State University, USA $^3$National Radio Astronomy Observatory, USA $^4$Dept. of Physics \& Astronomy, University of California, Los Angeles, CA, USA}

\pubyear{2016}
\volume{322}  
\jname{The Multi-Messenger Astrophysics of the Galactic Centre}
\editors{R. Crocker, S. N. Longmore \& G. Bicknell}
\begin{document}

\maketitle

\begin{abstract}
We present \am~and \rrltwo~VLA observations of the Radio Arc region, including the \scloud~and \gcloud~\mc s. These observations suggest the two velocity components of \scloud~are physically connected in the south. Additional ATCA observations suggest this connection is due to an expanding shell in the molecular gas, with the centroid located near the \qc. The \gcloud~molecular cloud has little radio continuum, strong molecular emission, and abundant \meth~masers, similar to a nearby \mc~with no star formation: \brick. These features detected in \gcloud~suggest dense molecular gas with no signs of current star formation.
\end{abstract}
\firstsection 
\section{Introduction}

Galactic centre (\gc) molecular clouds are suggested to follow `orbital streams' that are precessing, connected chains of molecular gas (\cite[Kruijssen {\it et al.} 2015]{Kru15}). 
\cite[Kruijssen {\it et al.} (2015, and in this volume)]{Kru15} recently proposed an open orbit model that is segmented into four orbital streams. However, modeling these orbital streams is difficult due to several complicated kinematic regions. The complexity of these regions is attributed to multiple streams overlapping along our line of sight (\cite[Kruijssen {\it et al.} 2015]{Kru15}; Figure 4).

{\underline{\it The Radio Arc:}} This region has complex kinematics. The \mc~at this location, \scloud~(\cite[Serabyn \& Guesten 1991]{SG91}), has two velocity components, that are located at 25 \& 80 \kms. \cite[Kruijssen {\it et al.} (2015, and in this volume)]{Kru15} argue that these two components to being on the near side (80 \kms; stream 1) and far side (25 \kms; stream 3) of the \gc. 
Adjacent to \scloud~is \gcloud, which is the core of a larger, dense \mc~(\cite[Tsuboi {\it et al.} 1997 Handa {\it et al.} 2006]{Tsuboi97, Handa06}). The orbital model by \cite[Kruijssen {\it et al.} (2015, and in this volume)]{Kru15} suggests that this cloud is near pericentre on the near side of the \gc~(stream 1).

\firstsection
\section{Observations}

We recently conducted sensitive continuum and spectral line observations with the VLA (\am, \meth, \rrl), of \scloud~and \gcloud, and present the results here. We compare these results with lower resolution ATCA data (\am) from Ott et al. (in this volume). These observations will be used to analyze the kinematic properties of the molecular gas (as traced by the \am~(3,3) transition, hereafter `molecular gas') and ionized gas (as traced by the combined \rrltwo~transitions, hereafter `ionized gas').

\firstsection
\section{Results and Discussion}

\begin{figure}[t]
 \vspace*{-1.0 cm}
\begin{center}
 \includegraphics[width=5.25in]{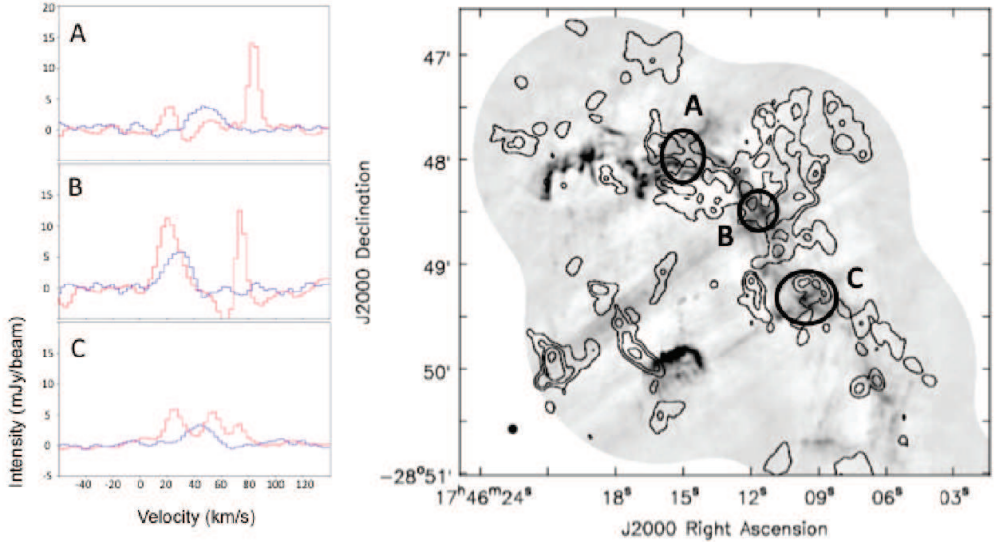} 
 \vspace*{-2.0 cm}
 \caption{{\it left}: three spectral panels of \am~(red; 2$-$3 components) and \rrltwo~(blue; single component) from the regions (A$-$C) in ({\it right}). {\it right}: 24.5 GHz continuum (greyscale) with contours of \am~emission at 5, 10, 20, \&~50 $\times$ 3 m\jyb~(rms level). The ionized gas emission (not shown) closely follows the continuum emission. }
   \label{fig1}
\end{center}
\end{figure}

{\underline{\it Connection between ionized and \amm~in M0.20-0.033}}: Our VLA observations also show two components in the \amm. Figure \ref{fig1} (right) shows the distribution of \amm~relative to the 24.5 GHz continuum emission. The spectral distribution of molecular and ionized gas for three selected regions (labeled A$-$C) are shown in Figure \ref{fig1} (left). In all three regions, only one ionized gas component is detected. This ionized gas component shows a strong correlation with the \low~component at position (B). The two northern regions (A \& B) show two molecular components separated in velocity space. The most southern region (C) has a third molecular component, located at intermediate velocities (\til50 \kms). The presence of molecular emission at these intermediate velocities suggests that the 25 \&~\high~components in \scloud~are physically connected in the south via this third velocity component.

\begin{figure}[t]
\begin{center}
 \includegraphics[width=5.25in]{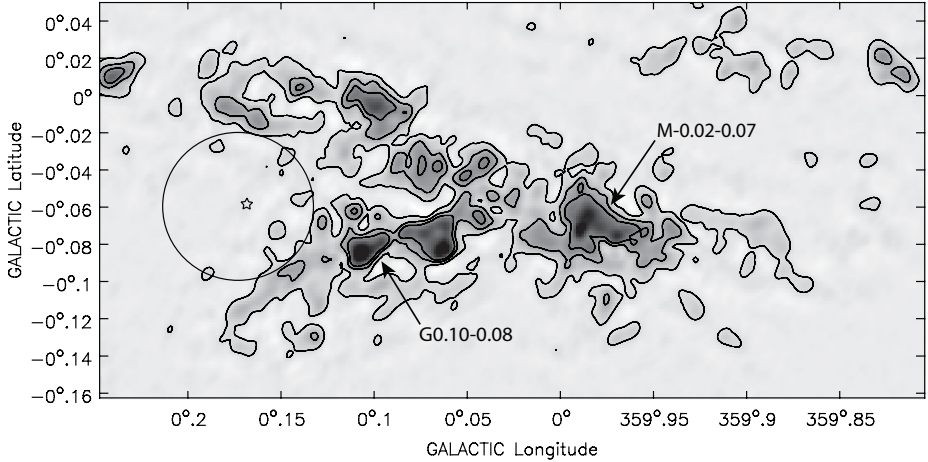} 
 \vspace*{-1.0 cm}
 \caption{\am~(3,3) emission at 50$-$60 \kms, using ATCA observations from the SWAG survey (17\arcsec~resolution). The black circle shows the location of our proposed expanding shell. The star denotes the location of the \qc. For more details see \butter. }
   \label{fig2}
\end{center}
\end{figure}

{\underline{\it Expanding shell in \amm}}: In order to confirm the connection between the two components in the southern region of \scloud~we examined ATCA data from the SWAG survey. These observations reveal a cavity in the molecular emission that is open on the eastern side (Figure \ref{fig2}). This cavity can be fit with a circle (hereafter, shell; Figure \ref{fig2}), that is centered at \ra=17\h46\m17.3\s, \dec=-28\degree49\arcmin00\arcsec, with a radius r=150\arcsec (\til6pc). An expanding shell produces an elliptical distribution in \pv~space, because the near side of the shell is blueshifted and the far side of the shell is redshifted. 
Our position-velocity analysis, using the ATCA data, suggests an expanding shell that has an expansion velocity of \til40 \kms, with a central velocity of \til51 \kms.
The \pv~distribution indicates that the \low~component (blueshifted emission) is on the near side of the shell and the \high~component (redshifted emission) is on the far side of the shell. 
\hi~absorption observations towards this region support such an arrangement (\cite[Lang {\it et al.} 2010]{lang10}). 
The central velocity of the shell at \til51 \kms~suggests that the shell is located on stream 1 (near side of the \gc; \cite[Kruijssen {\it et al.} 2015]{Kru15}). 

{\underline{\it Connection to \qc}}: 
The massive stellar \qc~(\cite[$N_{Lyc}$\til10$^{50.9}$ photons s$^{-1}$, $t_{age}$=4.8$\pm$1.1 Myr; Figer {\it et al.} 1999, Schneider {\it et al.} 2014]{schneider14}), is located adjacent to \scloud. 
The powerful \qc~is capable of ionizing the surrounding ISM, producing \sick~(\cite[Lang {\it et al.} 1997]{lang97}). 
The center of the expanding shell is located near (\til30\arcsec~in projection) the \qc~(star on Figure \ref{fig2}), suggesting that the shell may be produced by the \qc. 

The \cite[Kruijssen {\it et al.} (2015)]{Kru15} orbital model suggests that the \qc~is located on stream 1. As \scloud~is also suggested to be located on stream 1, the close proximity of the cluster and cloud indicates a possible interaction, as shown by \sick. 
Since our expanding shell has an estimated age of 1.5$\times$10$^{5}$ yr, assuming a constant expansion rate, this suggests that the interaction between the cluster and the cloud is relatively recent. 



{\underline{\it \gcloud: Another Brick?}}
The VLA observations of \gcloud~indicate that this \mc~has several interesting features:

\begin{figure}[t]
\begin{center}
 \includegraphics[width=5.25in]{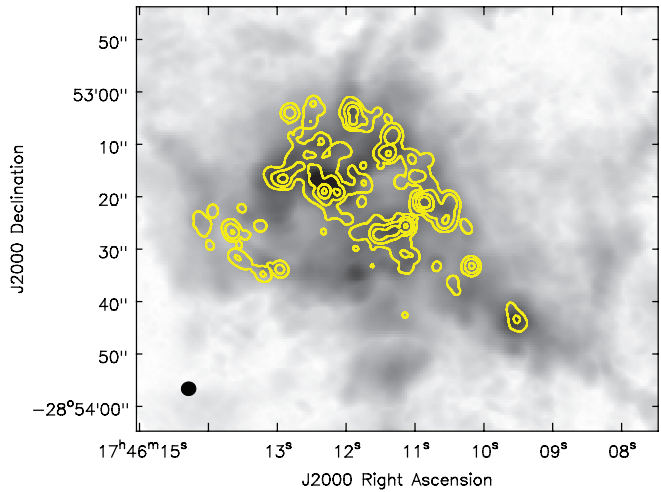} 
 \vspace*{-0.5 cm}
 \caption{maximum intensity \am~emission in \gcloud~(greyscale).  36 GHz \meth~maser contours at 2, 7, 15, and 50 per cent the peak emission (45.7 \jyb).  }
   \label{fig4}
\end{center}
\end{figure}

$\bullet$ {\bf Weak Radio Continuum}: \gcloud~has very little radio continuum at 24.5 GHz and no detected ionized gas emission. There is a very faint filamentary streak at 24.5 GHz, near the right edge of the cloud (\ra=17\h46\m07\s, \dec=$-$28\degree53\arcmin20\arcsec), that is the only emission detected above 5$\sigma$. This lack of radio continuum and ionized gas would suggest no embedded star formation in \gcloud. Since there is, however, slight external ionization near the right edge of the cloud, this would suggest ionization from an outside source. 

$\bullet$ {\bf Bright \am~emission}: The molecular emission in \gcloud~is bright, compared to \scloud. Figure \ref{fig4} (greyscale) shows the distribution of the molecular emission, which is abundant throughout the cloud. There are several bright, compact knots of molecular emission distributed throughout \gcloud. The morphology of these compact knots of emission is similar to those detected in \brick~(\cite[Mills {\it et al.} 2015]{mills15}).

$\bullet$ {\bf Kinematics}: The \amm~in \gcloud~ranges from \til30$-$70 \kms, and peaks in intensity at \til55 \kms. The velocity distribution shows a slight gradient (\til10 \kmsp), in the direction of positive galactic latitude. This gradient is consistent with other \gc~molecular clouds. 

$\bullet$ {\bf Abundant \meth~masers}: We have detected 64 compact 36 GHz \meth~sources (hereafter, \meth~masers) that have a brightness temperatures above 400 K (\cite[Mills {\it et al.} 2015; \butter]{mills15}). These \meth~masers follow the \amm~emission fairly well (Figure \ref{fig4}). 
The \meth~masers in \gcloud~are clustered towards the center of the cloud, where the molecular emission is the strongest. 
The clustering of \meth~masers detected in \gcloud~is similar to the `bar' of masers seen in the southern region of \brick~(\cite[Mills {\it et al.} 2015]{mills15}). The 36 GHz \meth~maser is a class I maser and therefore is a known shock tracer. Thus, the particularly high abundance of \meth~masers suggest shocked gas towards the core of \gcloud. 
 
Our VLA study suggests that \gcloud~is a dense, compact \mc~that shows no signs of current star formation.
The molecular characteristics of \gcloud~are similar to those seen in \brick~(\cite[also known as the `Brick'; Mills {\it et al.} 2015]{mills15}). \brick~is also suggested to be a quiescent, dense \mc~that is not currently forming stars (\cite[Longmore {\it et al.} 2012; Mills {\it et al.} 2015]{longmore12, mills15}). 
\firstsection

\end{document}